\documentclass[doublecol]{epl2}

\usepackage{amsmath}
\usepackage{graphicx}
\usepackage{amssymb}
\usepackage{bm}
\usepackage{longtable}
\usepackage{color}
\usepackage{epstopdf}
\usepackage{hyperref}

\newcommand{\vn}[1]{\mathbf{ #1}}
\newcommand{\mean}[1]{\left\langle {#1} \right\rangle }

\title{Temperature chaos in 3D Ising Spin Glasses is driven by rare events}
\author{L.A.~Fernandez\inst{1,2}, V. Martin-Mayor\inst{1,2}, G. Parisi\inst{3,4}, B. Seoane\inst{3,2}}
\shortauthor{L.A.~Fernandez, V. Martin-Mayor, G. Parisi and B. Seoane}
\institute{
\inst{1} Departamento de F\'\i{}sica Te\'orica I, 
  Universidad Complutense, 28040 Madrid, Spain.\\
\inst{2} Instituto de Biocomputaci\'on and
  F\'{\i}sica de Sistemas Complejos (BIFI), 50009 Zaragoza, Spain.\\
\inst{3} Dipartimento di Fisica, Sapienza Universit\`a di Roma.\\
\inst{4} INFN, Sezione di Roma I, IPFC-CNR, P.le A. Moro 2, I-00185 Roma, Italy.\\
}

\pacs{75.50.Lk}{Spin glasses and other random magnets}
\pacs{75.40.Mg}{Numerical simulation studies}

\abstract { Temperature chaos has often been reported in literature as a
  rare-event driven phenomenon. However, this fact has always been ignored in
  the data analysis, thus erasing the signal of the chaotic behavior (still
  rare in the sizes achieved) and leading to an overall picture of a weak and
  gradual phenomenon.  On the contrary, our analysis relies on a
  large-deviations functional that allows to discuss the size dependencies. In
  addition, we had at our disposal unprecedentedly large configurations
  equilibrated at low temperatures, thanks to the Janus computer. According to
  our results, when temperature chaos occurs its effects are strong and can be
  felt even at short distances.  }

\begin{document}
\maketitle 

Temperature chaos (TC) refers to the complete reorganization of the
equilibrium configurations by the slightest change in temperature.  This
effect was initially predicted in spin glasses
(SG)~\cite{mckay:82,bray:87b,banavar:87} but it is expected as well in other
glassy materials such as polymers~\cite{sales:02,dasilveira:04} or vortex
glasses~\cite{cieplak:93}.

An experimental measurement of TC is still missing. The main difficulty arises
from the nonequilibrium nature of the experimental glass: since chaos is an
\emph{equilibrium} property, it is not clear how to detect it in
nonequilibrium responses (such as the aging magnetic
susceptibility~\cite{vincent:97}, for instance). Nevertheless, TC is often
regarded as the origin of the anomalous response of glasses in temperature
cycles, and in particular, of the spectacular rejuvenation (and memory)
effects found in SG\cite{jonason:98} (however,
see~\cite{berthier:02,berthier:03} for a dissenting view).  Although memory
and rejuvenation have been also identified in colloids~\cite{ozon:03},
polymers~\cite{bellon:00,yardimci:03},
ferro-electrics~\cite{bouchaud:01b,mueller:04}, and in the ferromagnetic phase
of disordered magnetic alloys~\cite{vincent:00}, these are tiny effects as
compared to their SG counterpart.  Thus, experiments suggest that TC is
peculiar in SG.

Unfortunately, a rigorous theoretical description of TC has been achieved only
for directed polymers in random media in 1+1
dimensions~\cite{sales:02,dasilveira:04}. Coming to SG, the analytical work
has been mostly concerned with mean-field (MF) approximations. Some accidental
cancellations make TC anomalously weak in the Sherrington-Kirkpatrick
model~\cite{sherrington:75} (the standard model in MF approximations), which
favored a long controversy about its very
existence~\cite{kondor:89,kondor:93,billoire:00,rizzo:01,mulet:01,billoire:02,krzakala:02,rizzo:03}. Only
recently the question has been answered in the positive, by the explicit
computation of a large-deviation functional (i.e. the free-energy cost of
constraining a SG to have similar spin configurations at two temperatures
below the critical one, $T_1,T_2<T_\mathrm{c}$)~\cite{rizzo:03,parisi:10}. A
large-deviation functional will play as well a crucial role in this work.

Besides, the theoretical work in non MF models is restricted to
\emph{equilibrium} numerical
simulations~\cite{ney-nifle:97,ney-nifle:98,krzakala:04,sasaki:05,katzgraber:07}.
Data were analyzed using a scaling picture (valid for polymers), in which a
characteristic length-scale should appear $\xi_\mathrm{C}(T_1,T_2)$ in the
comparison of the system at two temperatures
$T_1,T_2<T_\mathrm{c}$~\cite{fisher:86,bray:87}.  Spin configurations at
temperatures $T_1$ and $T_2$ would be similar (different), if compared on
length-scales smaller (larger) than $\xi_\mathrm{C}(T_1,T_2)$. The chaos
length should diverge when $T_1$ approaches $T_2$ as $\xi_\mathrm{C}\propto
|T_2-T_1|^{-1/\zeta}$. The numerical evidence for this picture is rather weak,
which has been attributed to a large $\xi_\mathrm{C}(T_1,T_2)$, comparable or
larger than the simulated system sizes~\cite{aspelmeier:02}. Overall, the
emerging picture is that of a gradual and extremely weak phenomenon.

However, the situation is murkier than suggested by the scaling picture. In
the weak chaos scenario~\cite{sales:02} TC is almost absent in small systems
with the exception of very rare samples which are dramatically affected by
small temperature changes.  Weak chaos in SG has been reported in
numerics~\cite{katzgraber:07}, but a quantitative treatment  in the infinite volume is
lacking. Furthermore, we show below that the standard statistical analysis
sweeps under the rug these very few chaotic samples.

Here, TC in three-dimensional SG is
\emph{quantitatively} treated as a rare-event driven phenomenon.  When TC
occurs its effects are strong and can be felt even at the shortest length
scales (which contradicts the common belief of a very long chaotic length).
Finite-size effects are central to our approach because finite-size/time
scaling is our only bridge between theoretical equilibrium computations
(obtained for finite \emph{sizes}), and the nonequilibrium responses accessible
to experiment (at
finite~\emph{times})~\cite{franz:98,janus:10,janus:10b,barrat:01}. We achieve
here the characterization of equilibrium TC and of its system-size dependence,
thus paving the way for the analysis of temperature cycles.  Data suggest that
naive dimensional analysis breaks down: the chaotic-length diverges with
system size $L$ as $\xi_{\text{C}}(T_1,T_2)\propto L^{a}\,,\ a\approx 0.4$
(hence, $1\ll\xi_{\text{C}}(T_1,T_2)\ll L$ for large $L$). Two ingredients
were crucial in this work. First, the Janus computer~\cite{janus:08,janus:09}
gave us access to unprecedentedly large configurations, well thermalized up to
very low temperatures~\cite{janus:10,janus:10b}. Second, we introduce new
tools of statistical analysis, based on a large-deviations functional.

In this work, we reanalyze the Janus' equilibrium spin
configurations already used in ~\cite{janus:10,janus:10b} for the $D\!=\!3$
Edwards-Anderson model~\cite{edwards:75,edwards:76}. We consider Ising spins
$s_{\vn{x}}=\pm 1$, each placed in the $V=L^D$ nodes $\vn{x}$ of a cubic
lattice of linear size $L$, with periodic boundary conditions. The
interaction is restricted to lattice nearest neighbors. The coupling
constants $J_{\vn{x},\vn{y}}=\pm 1$ are chosen with $50\%$
probability. This model undergoes a SG transition at
$T_\mathrm{c}=1.109(10)$~\cite{hasenbusch:08b}. We study $4000$
realizations of disorder, named {\em samples}, for $L=8,\ 12,\ 16$ and
$24$ ($1000$ samples for $L=32$). The minimal temperature in the
Parallel Tempering simulation increased with $L$ (for $L=32$ it was
$T_\mathrm{min}= 0.7026$).

Given the aforementioned difficulties on numerical investigations, it is
crucial to chose wisely the quantities to be studied. Analytical 
approaches~\cite{rizzo:03,parisi:10} suggest the study of the probability
distribution function of the overlap between the spin configurations at temperatures $T_1$ and $T_2$,
\begin{equation}\label{eq:overlap}
q^{}_{T_1,T_2}=\frac{1}{V}\sum_{\vn{x}}   q_{\vn{x}}^{T_1,T_2}\,,\quad \text{with}\quad  q_{\vn{x}}^{T_1,T_2}= s_{\vn{x}}^{T_1} s_{\vn{x}}^{T_2}\,,
\end{equation}
or alternatively, a similar extension of the link-overlap,
\begin{equation}\label{eq:qlink}
Q^{\mathrm{link}}_{T_1,T_2}=\frac{1}{3V}\sum_{\Vert\vn{x}-\vn{y}\Vert=1}  q_{\vn{x}}^{T_1,T_2} q_{\vn{y}}^{T_1,T_2}\,,
\end{equation}
where the summation is restricted to lattice nearest-neighbors (it has been
argued that the link-overlap is the relevant order parameter below the upper
critical dimension~\cite{contucci:03,contucci:06}).

However, within the reachable system sizes, the spin overlap is not up to the
task for numerical simulations~\cite{billoire:02}. Rather, finite-size effects
are expected to be partly absorbed by a slight-modification, the {\em chaotic
  parameter}~\cite{ney-nifle:97}:
\begin{equation} \label{eq:X12}
X^J_{T_1,T_2}=\mean{q^2_{T_1,T_2}}_J\big/\big(\mean{q^2_{T_1,T_1}}_J\mean{q^2_{T_2,T_2}}_J\big)^{1/2}\,,
\end{equation}
Here, $\mean{\cdot}_J$ refers to thermal-averages within the same sample.
Note that $0\!<\! X^J_{T_1,T_2}\!\leq\!  1$. In fact, $X^J_{T_1,T_2}$ is
similar to a correlation parameter ($X^J_{T_1,T_2}\!=\!1$ means that, for that
particular sample, typical spin configurations at $T_1$ and $T_2$ are
indistinguishable, while $X^J_{T_1,T_2}\!=\!0$ indicates extreme
chaos). However, the standard analysis of $X^J_{T_1,T_2}$ (wrongly) concluded
that chaos was very weak.  For later reference, we remark that the chaotic
parameter can be generalized to the link overlap
\begin{equation} \label{eq:X12link}
X^{\mathrm{link},J}_{T_1,T_2}=\mean{Q^{\mathrm{link}}_{T_1,T_2}}_J\big/\Big(\mean{Q^{\mathrm{link}}_{T_1,T_1}}_J\mean{Q^{\mathrm{link}}_{T_2,T_2}}_J\Big)^{1/2}\,.
\end{equation}
We shall see that $X^{\mathrm{link},J}_{T_1,T_2}$ is just as informative as
$X^J_{T_1,T_2}$, as it could be expected from
replica-equivalence~\cite{parisi:00,contucci:06,janus:10b}.

We need some new insight to find a good
observable. Following~\cite{schulman:07,fernandez:06}, we will seek it in the
Monte Carlo dynamics, specifically in the temperature flow of the Parallel
Tempering~\cite{hukushima:96,Marinari:98b}. Indeed, if the equilibrium
configuration for two neighboring temperatures are too different (TC), a
bottleneck in the temperature random-walk should appear. Now, the performance
of Parallel Tempering deteriorates dramatically when the system size grows
from $L=8$ to $L=32$~\cite{janus:10}.  It follows that some strong form of TC
is waiting to be unveiled.

\begin{figure}[h]
\includegraphics[angle=270,width=\columnwidth]{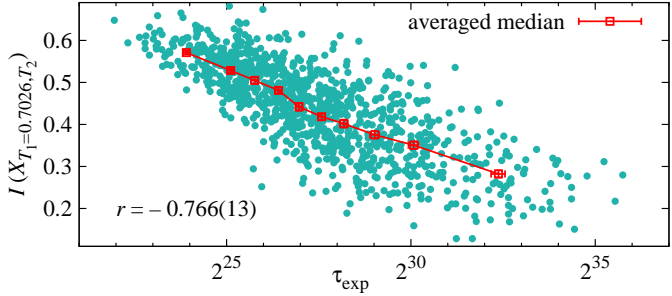}
\caption{(Color online) Seeking clues about TC on the parallel tempering
  autocorrelation time $\tau_\mathrm{exp}$~\cite{janus:10}. For each sample we
  show $I$ versus $\log{\tau_\mathrm{exp}}$, where
  $I\!=\!\int_{T_1}^{T_{\text{max}}} X^J_{T_1,T_2}\ \mathrm{d}T_2$ and
  $X^J_{T_1,T_2}$ is defined in Eq. \eqref{eq:X12}
  [$T_1\!=\!T_{\text{min}}\!=\!0.7026$, $T_\mathrm{max}\!=\! 1.549$, data for
    $L\!=\!32$]. The correlation parameter $r$ is computed with
  $\log\tau_\mathrm{exp}$ due to the wild sample to sample fluctuations of
  $\tau_\mathrm{exp}$. The red line was computed by a delicate procedure: we
  ordered the samples by increasing $\log{\tau_\mathrm{exp}}$ and made groups
  of 100 consecutive samples; within each group, medians were computed (errors
  from bootstrap).}
\label{fig:ob_vs_logtau}
\end{figure}

The temperature-flow dynamics is characterized by its exponential
autocorrelation time, $\tau_\mathrm{exp}$~\cite{sokal:97,janus:10}. Note that
$\tau_\mathrm{exp}$ is potentially unbounded, while $X^J_{T_1,T_2}$ is limited
within 0 and 1. Hence, $\tau_\mathrm{exp}$ should provide a clear flag
signaling those samples that suffer strong chaos. It follows that the quantity
that better correlates with $\log\tau_\mathrm{exp}$ will also be the most
informative about chaos.  After some failures~\cite{seoane-tesis}, we found
strong anticorrelation with the integral of $X^J_{T_1,T_2}$ with $T_2$,
$I=\int_{T_1}^{T_{\text{max}}} X^J_{T_1,T_2}\ \mathrm{d}T_2$, see
Fig. \ref{fig:ob_vs_logtau}. If $X^J_{T_1,T_2}$ suffers a sharp drop at low
$T_2$ (thus making the integral $I$ small), the temperature flow in that
sample is likely to get stuck. These samples are completely neglected
by the standard statistical analysis.

\begin{figure}[h]
\includegraphics[angle=270,width=\columnwidth]{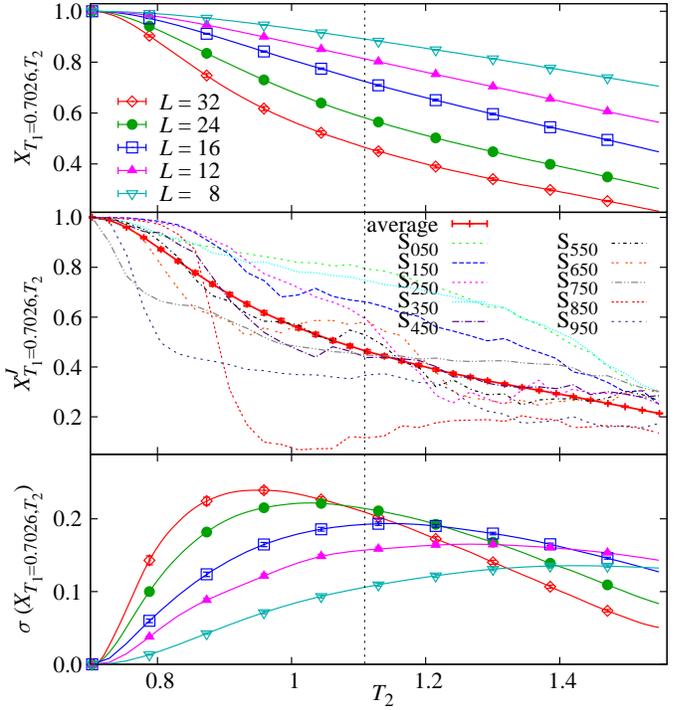}
\caption{(Color online) Different views on $X^J_{T_1,T_2}$,
  Eq.~\eqref{eq:X12}, as function of $T_2$ ($T_1\!=\!0.7026$, the vertical
  line is $T_2\!=\!T_\mathrm{c}$). {\bf (Top)} For all our system sizes,
  sample-average of $X^J_{T_1,T_2}$. {\bf (Center)} For $L=32$, we show
  $X^J_{T_1,T_2}$ for ten samples evenly spaced on a list of growing
  $\tau_\mathrm{exp}$, recall Fig~\ref{fig:ob_vs_logtau}. {\bf (Bottom)} For
  all our system sizes, we show the dispersion (i.e. square root of variance
  over the samples) of $X^J_{T_1,T_2}$.}
\label{fig:x12_y_var}
\end{figure}

Fig. \ref{fig:x12_y_var} describes the necessary change of paradigm.
The top panel shows the standard average over the samples of
$X^J_{T_1,T_2}$, as a function of $T_2$.  In agreement with previous
work~\cite{ney-nifle:97,ney-nifle:98,katzgraber:07}, our simulated
sizes are painfully away from the large-$L$ limit, where the average
of $X^J_{T_1,T_2}$ should vanish if $T_2\neq T_1$.  Instead, our
curves are smooth and cross $T_\mathrm{c}$ without qualitative
changes.  Yet, the behavior of individual samples is quite different,
Fig.~\ref{fig:x12_y_var}---center. At well defined temperatures $T_2$,
$X^J_{T_1,T_2}$ falls abruptly for some samples. This we name {\em
  chaotic event\/}.  The temperature at which these events occur is
random (many samples do not suffer any). In fact, as $L$ grows, the
sample dispersion of $X^J_{T_1,T_2}$ reaches a maximum in the SG
phase. This is the first time that a clear difference between the
paramagnetic and the SG phase is observed when studying
$X^J_{T_1,T_2}$.  We conclude that the full probability distribution
of $X^J_{T_1,T_2}$ should be studied.

\begin{figure}[h]
\includegraphics[angle=270,width=\columnwidth]{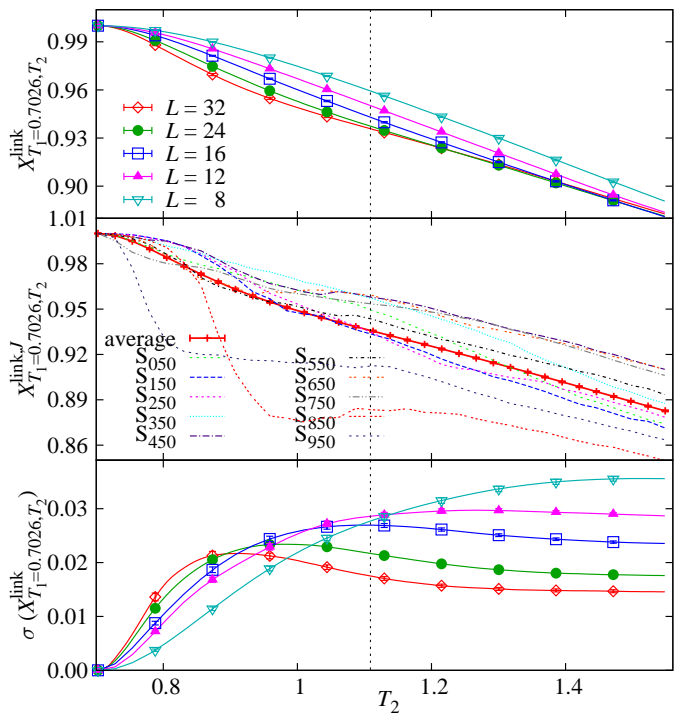}
\caption{(Color online) Same analysis as in Fig.~\ref{fig:x12_y_var}, carried
  out with the link-overlap $X^{\mathrm{link},J}_{T_1,T_2}$, see Eq.~\eqref{eq:X12link}.}
\label{fig:x12link_y_var}
\end{figure}

A natural question arises at this point. The link overlap carries information
only at distance one, supposedly much smaller than the chaos length
$\xi_\mathrm{C}(T_1,T_2)$~\cite{aspelmeier:02}. Can it carry as much
information as the spin-overlap, as expected from overlap equivalence
considerations? The answer is \emph{yes\/}, see Fig.~\ref{fig:x12link_y_var}
where we compare the spin and link chaotic parameters. Indeed, the chaotic
events observed in certain samples in Fig.~\ref{fig:x12_y_var}---center,
appear just as clearly for the very same samples in
Fig.~\ref{fig:x12link_y_var}---center. We will come back to this point below
when we discuss the spatial correlation function (in fact, the link overlap is
a correlation function at distance one).

\begin{figure}
\includegraphics[angle=270,width=\columnwidth]{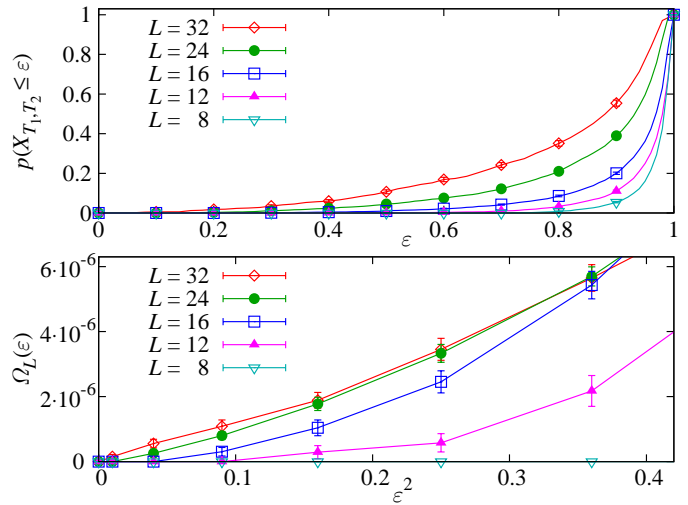}
\caption{(Color online) {\bf (Top)} Probability of finding $X^J_{T_1,T_2}\le
  \varepsilon$ as a function of $\varepsilon$, for temperatures $T_1= 0.7026$
  and $T_2=0.83923$ and all our system sizes.  {\bf (Bottom)} Large-deviation
  potential, Eq.~\eqref{eq:potential}, as a function of $\varepsilon^2$, from
  data in the top panel (for other $T_1$,$T_2$,
  see~\cite{seoane-tesis}). Chaos is absent for $L=8$, since
  $\varOmega_{T_1,T_2}$ is essentially zero. $\varOmega_{T_1,T_2}$ increases
  as $L$ grows, reaching a limiting behavior for $L=24$. }
\label{fig:potencial}
\end{figure}

We now consider the probability distribution of $X^J_{T_1,T_2}$. The fraction
of samples that suffer a chaotic event for any pair of temperatures $T_1,T_2$
($T_1<T_2<T_\mathrm{c}$) increases with $L$, see
Fig. \ref{fig:potencial}--top.  The statement is made quantitative by
introducing a large-deviation potential
$\varOmega^L_{T_1,T_2}(\varepsilon)$~\cite{rizzo:03,parisi:10}:
\begin{equation}\label{eq:potential}
\text{Probability}[X^J_{T_1,T_2}> \varepsilon]=\mathrm{e}^{-L^D\varOmega^L_{T_1,T_2}(\varepsilon)}\,.
\end{equation}
Several comments are in order:
\begin{itemize}
\item A large-deviation potential is useful only if
  $\varOmega^L_{T_1,T_2}(\varepsilon)$ becomes $L$-independent for moderate
  system sizes. This happens for $L=24$ and $32$ in
  Fig.~\ref{fig:potencial}--bottom (see~\cite{seoane-tesis} for other pairs of
  temperatures).
\item $\varOmega_{T_1,T_2}(\varepsilon)$ contains all the information on
  finite-size scaling. The values of $X^J_{T_1,T_2}$ that are likely on
  samples of size $L$ are such that $L^D
  \varOmega_{T_1,T_2}(X^J_{T_1,T_2})\sim 1$. This results on a power-law
  scaling, see Eq.~\eqref{eq:potential-scaling} below.
\item If $\varOmega_{T_1,T_2}(\varepsilon)$ remains positive for large $L$ and
  all $\varepsilon>0$, the probability of \emph{not} finding a chaotic sample
  is exponentially small in $L^{D}$.
\item $\varOmega_{T_1,T_2}(\varepsilon)$ can also be read as an
  $\varepsilon$-dependent length scale $\sim
  1/\varOmega^{1/D}_{T_1,T_2}(\varepsilon)$. This is \emph{not} the chaotic
  length of the scaling picture~\cite{fisher:86,bray:87,sales:02}, which is
  $\varepsilon$-\emph{independent.}
\end{itemize}

Chaos should weaken when $T_2$ approaches $T_1$. In fact,
MF~\cite{parisi:10} suggests the following scaling for the
large-deviation potential, in the limit $L\to\infty$, for small
$\varepsilon$ and $|T_1-T_2|$:
\begin{equation}\label{eq:potential-scaling}
\varOmega_{T_1,T_2}(\varepsilon)\propto |T_1 - T_2|^b \varepsilon^\beta\,.
\end{equation}
Our data for fixed $|T_1-T_2|$ suggest $\beta\approx 1.7$, see
Fig.~\ref{fig:potencial}--bottom and ~\cite{seoane-tesis}. Hence,
$\varOmega_{T_1,T_2}(\varepsilon)>0$ for all $\varepsilon>0$: the existence of
TC is established.

\begin{figure}
\includegraphics[angle=270,width=\columnwidth]{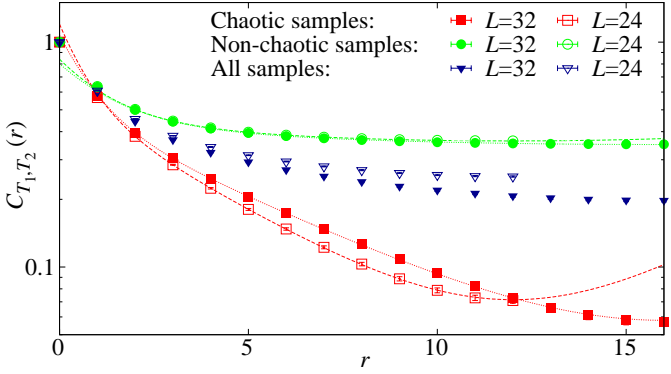}
\caption{(Color online) Spatial correlation function for $T_1=0.70260$ and
  $T_2=0.90318$, Eq.~\eqref{eq:cr}, as averaged over different sets of
  samples: all samples, non-chaotic samples
  ($X_{T_1,T_2}> 0.93$) and chaotic samples ($X_{T_1,T_2}\le
  0.33$). 
  Lines are fits to $\sum_{i=1,2}
  A_i\big(\mathrm{e}^{-x/\xi_i}+\mathrm{e}^{-(L-x)/\xi_i}\big)$. For each fit,
  the largest correlation lengths were $\xi_\mathrm{chaos}^{L=32}=5.69(2)$,
  $\xi_\mathrm{chaos}^{L=24}=4.447(15)$,
  $\xi_\mathrm{non-chaos}^{L=32}=23.7(7)$,
  $\xi_\mathrm{non-chaos}^{L=24}=18.9(3)$.}
\label{fig:cr}
\end{figure}

The exponent $\beta\approx 1.7$ in Eq.~\eqref{eq:potential-scaling} has some
consequences. To discuss them, it is enlightening to consider the spatial
correlation function for temperatures $T_1$ and $T_2$:
\begin{equation}\label{eq:cr}
C^J_{T_1,T_2}(r)=\frac{1}{3V}\sum_{\vn{r};|\vn{r}|=r}\sum_{\vn{x}}\mean{s^{T_1}_{\vn{x}}s^{T_2}_{\vn{x}}s^{T_1}_{\vn{x}+\vn{r}}s^{T_2}_{\vn{x}+\vn{r}}}_J\,.
\end{equation}
Now we average this function separately over the $10\%$ most (less) chaotic
samples.\footnote{In $L=32$, the $10\%$ most (less) chaotic samples are given
  by the samples that fulfill $X_{T_1,T_2}\le 0.33$ ($X_{T_1,T_2}> 0.93$). We
  keep the same $X_{T_1,T_2}$-thresholds for $L=24$.}, as well as the
totality of them, as shown in Fig.~\ref{fig:cr}. The behavior of the chaotic
samples is qualitatively different, as shown by the correlation-length
$\xi$.  On non-chaotic samples, we get $\xi_{\mathrm{NC}}\approx
L/2$ while, for the chaotic $L=32$ samples we obtain $\xi_{\mathrm{C}}\approx
6$ for $L=32$. As announced in the discussion of Eq.~\eqref{eq:potential},
an analysis based on a \emph{single} chaotic length~\cite{fisher:86,bray:87}
is incomplete.

The natural question, now, is how $\xi_{\mathrm{C}}$ behaves when $L$
grows \emph{for the typical samples} (i.e. those with $X_{T_1,T_2}^J$
such that $L^D \varOmega_{T_1,T_2}\sim
1$). Eq.~\eqref{eq:potential-scaling} provides a shocking answer:
$\xi_\mathrm{C}\sim L^{a}$, with $a=(\beta-1)/\beta$.\footnote{Indeed,
  barring normalizations, $\langle q_{T_1,T_2}^2\rangle_J$ is the
  space integral of the correlation function~\eqref{eq:cr}, recall
  Eq.~\eqref{eq:overlap}. It follows that $\langle
  q_{T_1,T_2}^2\rangle_J\propto (\xi_\mathrm{C}/L)^D$.  Since below
  $T_\mathrm{c}$, both $\langle q_{T_1,T_1}^2\rangle_J$ and $\langle
  q_{T_2,T_2}^2\rangle_J$ are of order one, also $X_{T_1,T_2}^J\sim
  (\xi_\mathrm{C}/L)^D$. Now, plug Eq.~\eqref{eq:potential-scaling}
  with $\beta\approx 2$ in Eq.~\eqref{eq:potential}. If the
  probability in Eq.~\eqref{eq:potential} is to remain of order one
  for large $L$, then $X_{T_1,T_2}^J\sim 1/L^{D/\beta}$ . But
  $X_{T_1,T_2}^J\sim (\xi_\mathrm{C}/L)^D$, so $\xi_\mathrm{C}\sim
  L^{a}$, with $a=(\beta-1)/\beta\approx 0.4$. The conclusion does not
  change if we consider an algebraically-decaying prefactor,
  $C^J_{T_1,T_2}(r)\sim \mathrm{exp}[-r/\xi_\mathrm{C}]/r^c$ for large
  $r$~\cite{berthier:02}. In such case, $\langle
  q_{T_1,T_2}^2\rangle_J\propto \xi_\mathrm{C}^{D-c}/L^D$ and
  $\xi_\mathrm{C}\sim L^{D(\beta-1)/(\beta(D-c))}$ (but our data
  suggest $c\approx 0$, $\beta\approx 1.7$). Furthermore, if for sizes
  $L>32$ one had $\beta=1$ (so that $\xi_\mathrm{C}$ would be finite
  for large $L$), then $\xi_\mathrm{C}\sim |T_1-T_2|^{-b/(D-c)}$.}
Our numerical results indicate that $a\approx 0.4$, so that $\xi_C$
diverges with $L$.  

Actually, only $\beta=1$ in Eq.~\eqref{eq:potential-scaling} would be
compatible with a finite $\xi_\mathrm{C}$.  However, we cannot neglect this
possibility. Indeed, the MF computation~\cite{parisi:10} warns us that
Eq.~\eqref{eq:potential-scaling} is probably only the leading order in an
expansion in $\epsilon$ and $|T_1-T_2|$. Further, subleading terms should be
expected. They would cause transient effects for small systems.~\footnote{In
  fact, in MF~\cite{parisi:10}, for small overlap and $|T_1-T_2|$, the
  large-deviations potential scales as
$$ \tilde\Omega^{\mathrm{mean-field}}\propto A q_{T_1,T_2}^2 |T_1-T_2|^3+B
  |q_{T_1,T_2}|^3 |T_1-T_2|^2\,,
$$ ($A$ and $B$ are constants). Either of the two terms can be dominant for
  some region of $q$, $N$ and $T_1-T_2$ ($N$ is the number of spins).  We now
  let $N$ grow at fixed $T_1-T_2$, and seek $q$ such that $\tilde \Omega\sim
  1/N$ [i.e.  probability of order one, see Eq.~\eqref{eq:potential}]. We
  realize that there is a crossover size $N^*\sim |T_1 -T_2|^{-5}$ such that
  $q^2\sim N^{-2/3}$ if $N\ll N^*$. On the other hand, if $N\gg N^*$, $q^2\sim
  N^{-1}$: the MF prediction for Eq.~\eqref{eq:potential-scaling} is
  $\beta=1$.}  In fact, the positive curvature that appears for large
$\epsilon$ in Fig.~\ref{fig:potencial}--bottom is probably due to these
contributions (subleading for small $\varepsilon$).  Thus, the
scaling displayed in Fig.~\ref{fig:potencial} could be still pre-asymptotic.

But, if the chaotic-length coming from the scaling picture cannot be given a
meaning (because the system sizes available to current simulation yield
$\beta\approx 1.7$, i.e an infinite $\xi_{\mathrm{C}}$), what is the chaos
exponent $\zeta$ computed in previous work~\cite{sasaki:05,katzgraber:07}?
(recall that, supposedly, $\xi_C\propto |T_2-T_1|^{-1/\zeta}$).  In fact, the
chaos exponent $\zeta$ was computed indirectly, through phenomenological
renormalization~\cite{banavar:87} ($\zeta\approx 1.07$ in
$D\!=\!3$~\cite{katzgraber:07} and $\zeta\approx 1.12$ in
$D\!=\!4$~\cite{sasaki:05}). We will now argue that this exponent is actually
$\zeta=D/b$ [$b$ is the temperature-difference exponent in
  Eq.~\eqref{eq:potential-scaling}]. So, $\zeta$ is unrelated to the chaos
length.

\begin{figure}
\includegraphics[angle=270,width=\columnwidth]{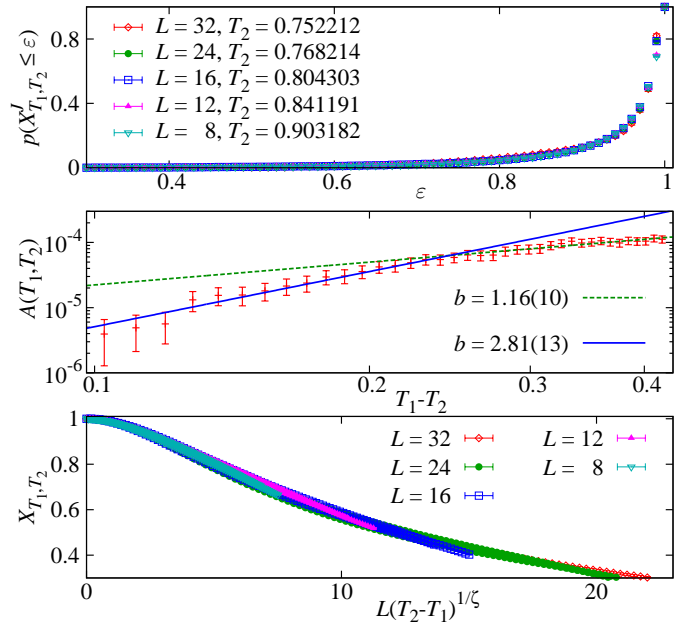}
\caption{(Color online) {\bf (Top)} Phenomenological renormalization: for each
  $L$, we seek $T_2(L)$ such that the distribution function
  $p\big(X^J_{T_1=0.7026,T_2(L)}\le \varepsilon\big)$ best resembles the $L=8$
  distribution for $T_1=0.7026$ and $T_2=0.90318$.  {\bf (Middle)} Computation
  of exponent $b$ in Eq.~\eqref{eq:potential-scaling}. First, we fit to
  $\varOmega^{L=32}_{T_1,T_2}(\varepsilon)=A(T_1,T_2)\varepsilon^2+B(T_1,T_2)\varepsilon^4$. We
  show $A(T_1,T_2)$ vs. $T_1-T_2$, and two power-law fits.  {\bf (Bottom)}
  Sample-averaged $X^J_{T_1,T_2}$ vs. $L(T_1-T_2)^{1/\zeta}$ using
  $\zeta=1.06$ (data for $T_1=0.7026$ and, when $L\leq 24$, also for
  $T_2=0.625$). Unfortunately, data collapse both for $T_2<T_\mathrm{c}$ and
  for $T_2$ in the paramagnetic phase (see also~\cite{ilker:13}). }
\label{fig:ajuste}
\end{figure}

Indeed, some reflection reveals that phenomenological
renormalization~\cite{katzgraber:07} can be cast as follows. For the purpose
of discussion, fix the lowest temperature $T_1$. Then, for each $L$, find a
$T_2(L)$ such that the probability distribution function for
$X_{T_1,T_2(L)}^J$, becomes $L$-independent, see
Fig.~\ref{fig:ajuste}--top. The exponent $\zeta$ of Ref.~\cite{katzgraber:07}
follows from $L\propto 1/|T_1 - T_2(L)|^{1/\zeta}$, see
Fig.~\ref{fig:ajuste}--bottom.\footnote{A fit to $T_1-T_2(L)\propto
  1/L^{\zeta}$ yields $\zeta=1.02(3)$ (for $L\leq 32$, $T_1=0.7026$ and
  $T_2(L=8)=0.90318$, with $\chi^2/\mathrm{dof}=3.57/3$) or $\zeta=1.07(2)$
  (for $L\leq 24$, $T_1=0.625$ and $T_2(L=8)=0.815$, with
  $\chi^2/\mathrm{dof}=1.77/2$). These results can be compared with
  $\zeta\approx 1.07$~\cite{katzgraber:07}.} But, combining
Eq.~\eqref{eq:potential} and~\eqref{eq:potential-scaling}, one realizes that
phenomenological renormalization amounts to the statement $L^{D} |T_1 -
T_2(L)|^b\sim 1$, which implies $\zeta=D/b$.

The computation of $\zeta$ from Eq.~\eqref{eq:potential-scaling} is reported
in Fig.~\ref{fig:ajuste}---center. There seems to be two scaling regimes.  For
small temperature differences, the regime where
Eq.~\eqref{eq:potential-scaling} applies, we find $b=2.81(13)$, or
$\zeta=1.07(5)$, in excellent agreement with~\cite{katzgraber:07}.  We note, however, that this $b$ value applies
only when $|T_1 - T_2(L)|<0.25$, when chaotic events are rather rare for our
system sizes. As anticipated in the Introduction, we see that the standard
statistical analysis is blind to TC.

It is time for some interpretation.  We have introduced the concept of a
chaotic event. Such chaotic events are pretty much similar to a "level
crossing" in Quantum Mechanics: at some specific temperature a new state
becomes favorable and takes over the old one. This is consistent with the fact
that chaotic events are noticeable even at distance one, see
Fig.~\ref{fig:x12link_y_var}. Now, our data suggest that, in the large $L$
limit and for any two temperatures $T_1$,$T_2< T_\mathrm{c}$ one should find a
chaotic event with probability one. Hence, given any pair of temperatures $T_a
< T_b$, one may consider a sequence $T_a<T_1$ < $T_2< T_3<\cdots<T_N<
T_b$. Chaotic events should appear in any subinterval $[T_i, T_{i+1}]$. It is
clear that a large amount of level crossings between temperatures $T_a<T_b$
can only result on the vanishing of the spin-overlap for both
temperatures. This picture bears some similarities to the findings within
Migdal-Kadanoff renormalization~\cite{ilker:13}, where a positive Lyapunov
exponents guarantee the vanishing of the spin-overlap (we remark that the
positive Lyapunov exponent appears within this approach for a large variety of
models, both regarding the spin-type and the coupling distribution).

In summary, we have \emph{quantitatively} characterized TC in the $D\!=\!3$
Ising SG as a rare-event driven phenomenon, inaccessible to the statistical
analysis employed in previous work. Instead, we perform a large-deviations
analysis of the Janus equilibrium configurations (remarkable both for system
sizes up to $L=32$ and for the low temperatures).  The large deviation
functional characterize the size-dependencies, a crucial step towards the
extension of time-length dictionaries for temperature-varying experimental
protocols.  A surprising outcome is that the chaotic length scales with system
size as $\xi_\mathrm{C} \!\propto\! L^a$, with $a\approx 0.4$: divergent in
the thermodynamic limit, yet much smaller than $L$. This duality will probably
be important to interpret the somehow contradictory memory and rejuvenation
effects.  In fact, although the $\xi_\mathrm{C}\!\propto\!L^a$ scaling follows
from a $L\to\infty$ extrapolation (which is tricky even in MF), we now know
that a meaningful comparison with experiments requires only an extrapolation
to $L\sim 110$ lattice spacings~\cite{janus:10}.

We are indebted with the Janus Collaboration for allowing us to analyze their
thermalized configurations. We thank particularly David Yllanes for
discussions and for assistance in the data analysis. We acknowledge support
from MINECO, Spain, through research contracts FIS2012-35719-C02 and from the
European Research Council (ERC) through grant agreement No. 247328.  B.S. was
supported by the FPU program (MECD, Spain).

\end{document}